\documentclass[10pt,journal,twocolumn]{IEEEtran}
\IEEEoverridecommandlockouts
\usepackage{cite}
\usepackage{amsmath,amssymb,amsfonts,booktabs}
\usepackage{graphicx}
\usepackage{textcomp}
\usepackage{xcolor}
\usepackage{enumerate}
\usepackage{epstopdf}
\usepackage{subcaption}
\usepackage{amsthm}

\usepackage{algorithm}
\usepackage{algorithmic}

\allowdisplaybreaks[4]

\begin{document}

\title{ \huge {QoS-Aware Integrated Sensing, Communication, and Control with Movable Antenna}}

\author{
\IEEEauthorblockN{  {
Yike Wang\IEEEauthorrefmark{1}\IEEEauthorrefmark{2},
Zhike Wu\IEEEauthorrefmark{3}\IEEEauthorrefmark{2},
Jiang Chen\IEEEauthorrefmark{4}\IEEEauthorrefmark{2},
Chunjie Wang\IEEEauthorrefmark{2}\IEEEauthorrefmark{5},
Xuhui Zhang\IEEEauthorrefmark{6},
and Yanyan Shen\IEEEauthorrefmark{2}} \\ }
\vspace{3.5pt}
\IEEEauthorblockA{ \small
\IEEEauthorrefmark{1}Wuhan Research Institute of Posts and Telecommunications, Wuhan, China\\
\IEEEauthorrefmark{2}Shenzhen Institutes of Advanced Technology, Chinese Academy of Sciences, Shenzhen, China\\
\IEEEauthorrefmark{3}University of Science and Technology of China, Hefei, China\\
\IEEEauthorrefmark{4}Southern University of Science and Technology, Shenzhen, China\\
\IEEEauthorrefmark{5}University of Chinese Academy of Sciences, Beijing, China\\
\IEEEauthorrefmark{6}Shenzhen Future Network of Intelligence Institute, The Chinese University of Hong Kong, Shenzhen, China\\
\vspace{-0.75em} 
}
\vspace{-18.5pt}
}

\maketitle
\begin{abstract}
Integrated sensing, communication, and control (ISCC) has emerged as a key enabler for low-altitude wireless networks with
enhanced adaptability through resource allocation co-design and intelligent environment awareness. However, dynamic interference and channel attenuation constrain the potential of the ISCC system.
To address this challenge, we propose a novel movable antenna-empowered ISCC system.
An achievable data rate maximization problem is formulated while guaranteeing the sensing and control quality-of-service (QoS) by optimizing the positions of the antennas and the beamforming strategy for communication, sensing, and control co-design.
An efficient alternating optimization (AO)-based algorithm is proposed to solve the highly coupled non-convex problem.
Numerical results demonstrate that the proposed AO-based algorithm achieves substantial gains in the achievable data rate and the control QoS compared with benchmark schemes.

\end{abstract}
\section{Introduction}
To bridge the coverage gap between conventional terrestrial cellular networks and high-altitude satellite systems, low-altitude wireless networks (LAWN) have emerged as a promising enabling infrastructure~\cite{yuan2025ground}. LAWNs operate in the airspace below $3,000$ meters and are supported by three core architectural planes: the data plane, the control plane, and the sensing plane. These planes work synergistically to deliver flexible, reconfigurable, and energy-efficient communication and sensing services, enabling diverse applications such as commercial drone delivery, emergency response, public safety, and large-scale environmental monitoring~\cite{liu2025movable}.
A key advantage of the LAWN is its collaborative sensing capability, which allows aerial and ground nodes to share environmental data in real time, achieving integrated air-ground information fusion~\cite{10693833}. To fully harness this potential, integrated sensing, communication, and control (ISCC) has been recently proposed~\cite{10907978}. The ISCC paradigm promotes close cooperation among the three functionalities through joint resource allocation, shared signal processing, and coordinated decision-making, thereby enhancing system intelligence, adaptability, and robustness. It has become a critical enabler for advanced applications including intelligent transportation, industrial automation, and integrated space-air-ground networks.

Nonetheless, practical implementation of the ISCC systems faces significant challenges. High demands on the number of antennas and the transmit power are exacerbated in high-frequency bands such as millimeter-wave and terahertz, where signals suffer from severe blockage and rapid attenuation due to the atmospheric absorption and limited diffraction, drastically reducing coverage. Furthermore, in obstructed environments, non-line-of-sight (NLoS) multi-hop paths often lead to severe path-loss and degraded beamforming gain, resulting in the poor communication rate, sensing and control performance, and widening the gap of the achievable service coverage, posing a major challenge for the design of balanced and efficient ISCC systems~\cite{10879807}.

To address these challenges, multi-antenna-enabled base stations (BSs) play a critical role in the ISCC-enabled systems. By leveraging spatial multiplexing, multi-antenna technology enhances spectral efficiency and increases spatial degrees of freedom (DoF) without additional spectrum usage~\cite{10379539}. Equipped with high energy and computational resources, the BSs can deploy large antenna arrays to exploit spatial DoF, thereby improving both communication reliability and sensing accuracy~\cite{11051123}. However, conventional systems face limitations such as signal blockage and multi-path fading, particularly in tracking dynamic target, where fixed antenna arrays struggle to maintain accurate beam alignment and continuous coverage~\cite{10354003, 10654366}.
In contrast, movable antennas (MAs) offer inherent spatial flexibility. By dynamically adjusting their positions and orientations, MAs can fully exploit continuous spatial DoF~\cite{10318061}, enabling enhanced spatial diversity, precise beam focusing, and improved resilience to fading. This leads to stronger link stability and faster responsiveness, making MAs a promising solution for dynamic ISCC environments~\cite{zhang2025movable}.

Based on the above observations and challenges, this paper proposes a novel ISCC system empowered by the MAs. 
We formualte a sum achievable user data rate maximization problem by jointly optimizing the positions of the MAs and the transmit beamforming at the BS, while satisfying the quality-of-service (QoS) requirements for sensing and control performance.
To address the non-convex optimization problem, we develop an efficient alternating optimization (AO)-based algorithm that iteratively optimize the MA positions and the transmit beamforming. Numerical results demonstrate the superior performance of the proposed AO-based algorithm compared to benchmark schemes, underscoring the significant potential of the MA-empowered ISCC systems for next-generation wireless networks.

\begin{figure}[ht]
    \centering \includegraphics[width=0.45\textwidth, height=0.27\textheight]{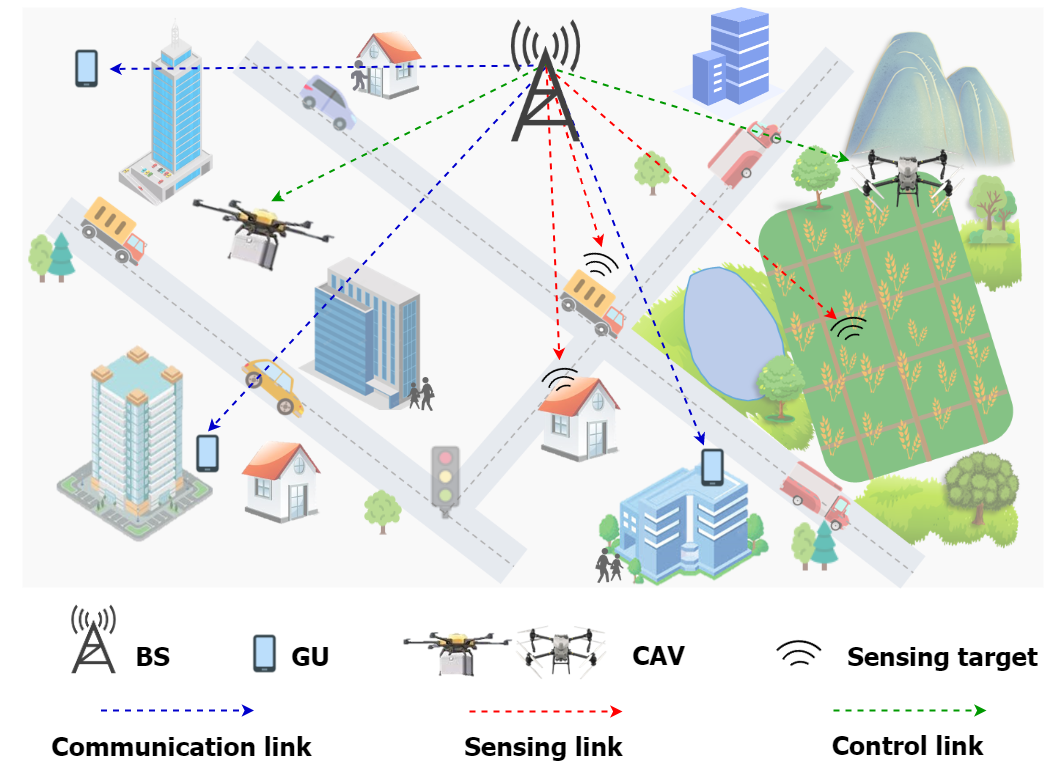}
    \caption{Illustration of a cooperative ISCC network deployed across urban and rural environments, featuring a multifunctional BS equipped with MAs.}
    \label{fig:system_img}
\end{figure}

\section{System Model}
As illustrated in Fig.~\ref{fig:system_img}, we consider an ISCC system comprising a multi-functional BS equipped with $M$ vertically deployed MAs, serving $J$ controlled aerial vehicles (CAVs), $N$ ground users (GUs), and $K$ sensing targets. The corresponding index sets are denoted as $\mathcal{M} = \{1, \dots, M\}$, $\mathcal{J} = \{1, \dots, J\}$, $\mathcal{N} = \{1, \dots, N\}$, and $\mathcal{K} = \{1, \dots, K\}$, respectively.  
The BS simultaneously transmits communication signals to the GUs, control signals to the CAVs, and dedicated sensing signals for sensing the targets, enabling fully integrated ISCC operations.  

We adopt a three-dimensional Cartesian coordinate system to model the spatial layout of the system. The BS is located at $\mathbf{q}_{\rm{BS}} = (\text{x}_{\rm{BS}}, \text{y}_{\rm{BS}}, \text{z}_{\rm{BS}})$. The positions of the $j$-th CAV, $n$-th GU, and $k$-th sensing target are denoted by $\mathbf{q}_j = ({\text{x}}_j, \text{y}_j, \text{z}_j)$, $\mathbf{q}_n = (\text{x}_n, \text{y}_n, \text{z}_n)$, and $\mathbf{q}_k = (\text{x}_k, \text{y}_k, \text{z}_k)$, respectively. The position of the $m$-th MA is represented as a two-dimensional coordinate vector $\mathbf{s}_m = (\text{x}_m, \text{y}_m)^{\mathsf{T}} \in \mathcal{C}_t$, where $\text{x}_m$ and $\text{y}_m$ denote its horizontal and vertical coordinates. Without loss of generality, the feasible movement region $\mathcal{C}_t$ is modeled as a square area of size $D \times D$, within which each MA can be dynamically repositioned to optimize the system performance.

\subsection{Communication Model}
The array response vector corresponding to the angle-of-departure (AoD) of the communication link transmitted from the BS to the $n$-th GU is given by
\begin{equation}
\mathbf{\bar{a}}_n = \left( {\mathrm{e}}^{ {\mathrm{j}} \frac{2\pi}{\lambda} \mathbf{p}_n \mathbf{s}_1}, {\mathrm{e}}^{ {\mathrm{j}} \frac{2\pi}{\lambda} \mathbf{p}_n \mathbf{s}_2},
\ldots, 
{\mathrm{e}}^{ {\mathrm{j}} \frac{2\pi}{\lambda} \mathbf{p}_n \mathbf{s}_M}
\right)^{\mathsf{T}} , n \in \mathcal{N},
\end{equation}
where $\lambda$ is the signal wavelength. $\mathbf{p}_n = \left( \sin \theta_n \cos \phi_n,\cos \theta_n \right)$ denotes the virtual AoD vector between the BS and the \(n\)-th GU, where \(\theta_n\) and \(\phi_n\) denote the elevation and azimuth AoDs, respectively \cite{10354003}.

As a result, the channel vector $\mathbf{h}_n \in \mathbb{C}^{M \times 1}$ between the BS and the $n$-th GU is given by
\begin{equation}
\mathbf{h}_n = \beta_0 d_n^{-\alpha} \left( 
\sqrt{ \frac{K_n}{K_n + 1} } \mathbf{\bar{a}}_n 
+ \sqrt{ \frac{1}{K_n + 1} } \mathbf{\tilde{a}}_n 
\right), n \in \mathcal{N},
\end{equation}
where $\beta_0$ denotes the path-loss at a reference distance of $1\,\text{m}$, $\alpha$ is the path-loss exponent, and $d_n = \| \mathbf{q}_{\rm{BS}} - \mathbf{q}_n \|$ represents the Euclidean distance between the BS and the $n$-th GU. The parameter $K_n$ is the Rician factor, which characterizes the relative strength of the line-of-sight (LoS) component and NLoS component. The NLoS component $\mathbf{\tilde{a}}_n$ models small-scale fading and is assumed to follow a circularly symmetric complex Gaussian (CSCG) distribution, i.e., $\mathbf{\tilde{a}}_n \sim \mathcal{CN}(\mathbf{0}, \mathbf{I}_M)$.

Moreover, the transmitted signal from the BS is given by
\begin{equation}
\mathbf{x}_{\text{t}} = \sum_{n=1}^{N} \mathbf{w}_{n} s_{n}^{\mathrm{comm}} + \sum_{j=1}^{J} \mathbf{w}_{j} s_{j}^{\mathrm{control}} + \mathbf{s}_{0}, n \in \mathcal{N},\ j \in \mathcal{J},
\end{equation}
where $\mathbf{w}_{n}, \mathbf{w}_{j} \in \mathbb{C}^{M \times 1}$ denote the transmit beamforming vectors for the $n$-th GU and the $j$-th CAV, respectively. $s_{n}^{\mathrm{comm}}, s_{j}^{\mathrm{control}} \sim \mathcal{CN}(0,1)$ represent independent unit-power communication and control data symbols, which are assumed to follow the CSCG. The term $\mathbf{s}_{0} \in \mathbb{C}^{M \times 1}$ denotes the dedicated sensing signal, modeled as a zero-mean CSCG random vector with covariance matrix $\mathbf{R}_{0} = \mathbb{E} \left\{ \mathbf{s}_{0} \mathbf{s}_{0}^{\mathsf{H}} \right\} \succeq \mathbf{0}$.

The received signal-to-interference-plus-noise ratio (SINR) at the $n$-th GU is denoted by $\gamma_n$, and is given by
\begin{equation}
\gamma_n = \frac{|\mathbf{h}_n^{\mathsf{H}} \mathbf{w}_{n}|^2}
     {\sum_{i \ne n}^N |\mathbf{h}_n^{\mathsf{H}} \mathbf{w}_{i}|^2  
     + \sum_{j=1}^{J} |\mathbf{h}_n^{\mathsf{H}} \mathbf{w}_{j}|^2 
     + \mathbf{h}_n^{\mathsf{H}} \mathbf{R}_{0} \mathbf{h}_n
     + \sigma_n^2},
\end{equation}
where $\sigma_n^2$ denotes the additive white Gaussian noise (AWGN) power at the $n$-th GU.
Then, the achievable data rate of the $n$-th GU is given by
\begin{equation}
R_n = \log_2 \left( 1 + \gamma_n \right). \label{rate}
\end{equation}

\subsection{Sensing Model}
Without loss of generality, single-path model is adopted for sensing, as multi-hop reflected signals may suffer from severe path-loss. Therefore, we assume that only the LoS path from the BS to the target. Consequently, the array steering vector $\mathbf{\bar{a}}_k $ from the BS toward the $k$-th sensing target is given by
\begin{equation}
\mathbf{\bar{a}}_k =  \left( {\mathrm{e}}^{ {\mathrm{j}} \frac{2\pi}{\lambda} \mathbf{p}_k \mathbf{s}_1}, {\mathrm{e}}^{ {\mathrm{j}} \frac{2\pi}{\lambda} \mathbf{p}_k \mathbf{s}_2},
\dots, 
{\mathrm{e}}^{ {\mathrm{j}} \frac{2\pi}{\lambda} \mathbf{p}_k \mathbf{s}_M}
\right)^{\mathsf{T}} , k \in \mathcal{K},
\end{equation}
where $\mathbf{p}_k = \left( \sin \theta_k \cos\phi_k,\, \cos \phi_k \right)$ is the direction vector determined by the elevation angle $\theta_k$ and azimuth angle $\phi_k$.

To ensure adequate sensing QoS, the total beam-pattern gain toward the $k$-th target must exceed a predefined threshold. Then, the QoS constraint is expressed as
\begin{equation}
\mathbf{\bar{a}}_k^{\mathsf{H}} \left( 
\sum_{j=1}^{J} \mathbf{w}_j \mathbf{w}_j^{\mathsf{H}} + 
\sum_{n=1}^{N} \mathbf{w}_n \mathbf{w}_n^{\mathsf{H}} + 
\mathbf{R}_0 
\right) \mathbf{\bar{a}}_k \geq d_k^{2} \Gamma, k \in \mathcal{K},
\end{equation}
where $d_k = \| \mathbf{q}_{\rm{BS}} - \mathbf{q}_k \|$ is the distance between the BS and the $k$-th target, $\Gamma$ denotes the minimum QoS requirement of the beam-pattern gain towards targets.

\subsection{Control Model}
Similar to the communication links, the channel vector between the BS and the $j$-th CAV can be given by
\begin{equation}
\mathbf{h}_j = \beta_0 d_j^{-\alpha} \left( 
\sqrt{ \frac{K_j}{K_j + 1} } \mathbf{\bar{a}}_j 
+ \sqrt{ \frac{1}{K_j + 1} } \mathbf{\tilde{a}}_j 
\right), j \in \mathcal{J},
\end{equation}
where $d_j = \| \mathbf{q}_{\rm{BS}} - \mathbf{q}_j \|$ denotes the Euclidean distance between the BS and the $j$-th CAV. The NLoS component $\mathbf{\tilde{a}}_j \in \mathbb{C}^{M \times 1}$ is assumed to follow a CSCG distribution, i.e., $\mathbf{\tilde{a}}_j \sim \mathcal{CN}(\mathbf{0}, \mathbf{I}_M)$. The array steering vector $\mathbf{\bar{a}}_j$ for the $j$-th CAV is given by
\begin{equation}
\mathbf{\bar{a}}_j = \left( 
{\mathrm{e}}^{{\mathrm{j}} \frac{2\pi}{\lambda} \mathbf{p}_j \mathbf{s}_1},\,
{\mathrm{e}}^{{\mathrm{j}} \frac{2\pi}{\lambda} \mathbf{p}_j \mathbf{s}_2},\,
\ldots,\,
{\mathrm{e}}^{{\mathrm{j}} \frac{2\pi}{\lambda} \mathbf{p}_j \mathbf{s}_M}
\right)^{\mathsf{T}}, j \in \mathcal{J},
\end{equation}

Therefore, the received SINR at the $j$-th CAV is given by
\begin{equation}
\gamma_j = \frac{|\mathbf{h}_j^{\mathsf{H}} \mathbf{w}_{j}|^2}
     {\sum_{i \ne j}^{J} |\mathbf{h}_j^{\mathsf{H}} \mathbf{w}_{i}|^2 
     + \mathbf{h}_j^{\mathsf{H}} \mathbf{R}_0 \mathbf{h}_j 
     + \sum_{n=1}^{N} |\mathbf{h}_j^{\mathsf{H}} \mathbf{w}_{n}|^2 
     + \sigma_j^2},
\end{equation}
where $\sigma_j^2$ denotes the AWGN power at the $j$-th CAV. Then, the achievable data rate of the $j$-th CAV is given by
\begin{equation}
R_j =  \log_2 \left( 1 + \gamma_j \right), \quad j \in \mathcal{J}.
\label{control_rate}
\end{equation}

To enable reliable remote control and coordination of the CAVs, each CAV is modeled as a discrete-time linear time-invariant (LTI) dynamical system~\cite{704994,1099832}. The state evolution of the $j$-th CAV at time step $t$ is described by the state-space equation
\begin{equation}
\mathbf{x}_{j,t+1} = \mathbf{A}_j \mathbf{x}_{j,t} + \mathbf{B}_j \mathbf{z}_{j,t} + \mathbf{v}_{j,t}, \quad t = 1, 2, \ldots, T,
\end{equation}
where $\mathbf{x}_{j,t} \in \mathbb{R}^{n_1}$ denotes the state vector (e.g., position, velocity, and orientation), and $\mathbf{z}_{j,t} \in \mathbb{R}^{n_2}$ represents the control input vector. The process noise $\mathbf{v}_{j,t} \in \mathbb{R}^{n_1}$ is modeled as zero-mean white Gaussian noise with covariance matrix $\boldsymbol{\Sigma}_{v,j} \in \mathbb{R}^{n_1 \times n_1}$. $\mathbf{A}_j \in \mathbb{R}^{n_1 \times n_1}$ and $\mathbf{B}_j \in \mathbb{R}^{n_1 \times n_2}$ are pre-known and time-invariant, capturing the intrinsic dynamics of the $j$-th CAV.

The measurement at the BS is given by
\begin{equation}
\mathbf{y}_{j,t} = \mathbf{G}_j \mathbf{x}_{j,t} + \boldsymbol{\omega}_{j,t},
\end{equation}
where $\mathbf{y}_{j,t} \in \mathbb{R}^{n_3}$ is the measurement vector, $\mathbf{G}_j \in \mathbb{R}^{n_3 \times n_1}$ is the observation matrix, and $\boldsymbol{\omega}_{j,t} \in \mathbb{R}^{n_3}$ denotes the zero-mean white Gaussian measurement noise with covariance $\boldsymbol{\Sigma}_{\omega,j} \in \mathbb{R}^{n_3 \times n_3}$.

To assess control performance, the linear quadratic regulator (LQR) cost function is adopted~\cite{10460602}. The average LQR cost is defined as
\begin{align}
\mathrm{LQR}_{j,t} \triangleq & \ \mathbb{E} \bigg[ \sum_{i=1}^{t-1} \left( \mathbf{x}_{j,i}^{\mathsf{T}} \mathbf{Q}_j \mathbf{x}_{j,i} + \mathbf{u}_{j,i}^{\mathsf{T}} \mathbf{R}_j \mathbf{u}_{j,i} \right) \nonumber \\
& \quad + \mathbf{x}_{j,t}^{\mathsf{T}} \mathbf{Q}_{1,j} \mathbf{x}_{j,t} \bigg],
\end{align}
where $\mathbf{Q}_j, \mathbf{Q}_{1,j}, \mathbf{R}_j \succeq \mathbf{0}$ are symmetric weighting matrices that balance state deviations and control effort, respectively. 

As indicated in~\cite{8693967}, the achievable data rate must not be less than a minimum threshold to guarantee the QoS requirement, i.e., the system stability and the desired control performance, which can be expressed as
\begin{equation}
R_{j} \ge \eta_j + \frac{n_1}{2} \log \left( 1 + \frac{n_1 \left( \det(\mathbf{N}_j \mathbf{M}_j) \right)^{\frac{1}{n_1}}}{\bar{l}_j - \bar{l}_{j,\min}} \right), \quad j \in \mathcal{J},
\end{equation}
where $\eta_j \triangleq \log |\det(\mathbf{A}_j)|$ represents the intrinsic entropy rate associated with the stability of the $j$-th CAV, and $\bar{l}_j = \frac{1}{T} \mathrm{LQR}_{j,T}$ denotes the average LQR cost. $\mathbf{N}_j$ and $\mathbf{M}_j$ are given by
\begin{equation}
\mathbf{N}_j = \mathbf{A}_j \boldsymbol{\Sigma}_j \mathbf{A}_j^{\mathrm{T}} - \boldsymbol{\Sigma}_j + \boldsymbol{\Sigma}_{v,j},
\end{equation}
\begin{equation}
\mathbf{M}_j = \mathbf{S}_j \mathbf{B}_j \left( \mathbf{R}_j + \mathbf{B}_j^{\mathrm{T}} \mathbf{S}_j \mathbf{B}_j \right)^{-1} \mathbf{B}_j^{\mathrm{T}} \mathbf{S}_j,
\end{equation}
with $\mathbf{S}_j$ being the unique positive semidefinite solution to the discrete algebraic Riccati equation (DARE), and is given by
\begin{equation}
\mathbf{S}_j = \mathbf{Q}_j + \mathbf{A}_j^{\mathrm{T}} \left( \mathbf{S}_j - \mathbf{M}_j \right) \mathbf{A}_j.
\end{equation}
$\boldsymbol{\Sigma}_j$ denotes the steady-state covariance of the state estimation error, and can be expressed as
\begin{equation}
\boldsymbol{\Sigma}_j = \mathbf{P}_j - \mathbf{K}_j \left( \mathbf{G}_j \mathbf{P}_j \mathbf{G}_j^{\mathrm{T}} + \boldsymbol{\Sigma}_{\omega,j} \right) \mathbf{K}_j^{\mathrm{T}},
\end{equation}
where $\mathbf{P}_j$ is also the solution to a DARE, which is given by
\begin{align}
\mathbf{P}_j = \ & \mathbf{A}_j \mathbf{P}_j \mathbf{A}_j^{\mathrm{T}} - \mathbf{A}_j \mathbf{K}_j \left( \mathbf{G}_j \mathbf{P}_j \mathbf{G}_j^{\mathrm{T}} + \boldsymbol{\Sigma}_{\omega,j} \right) \mathbf{K}_j^{\mathrm{T}} \mathbf{A}_j^{\mathrm{T}} \nonumber \\
& + \boldsymbol{\Sigma}_{v,j},
\end{align}
and $\mathbf{K}_j$ denotes the steady-state Kalman gain, and is given by
\begin{equation}
\mathbf{K}_j = \mathbf{P}_j \mathbf{G}_j^{\mathrm{T}} \left( \mathbf{G}_j \mathbf{P}_j \mathbf{G}_j^{\mathrm{T}} + \boldsymbol{\Sigma}_{\omega,j} \right)^{-1}.
\end{equation}
Finally, $\bar{l}_{j,\min}$ denotes the minimum achievable LQR cost, which can be expressed as
\begin{equation}
\bar{l}_{j,\min} = \text{tr} \left( \boldsymbol{\Sigma}_{\omega,j} \mathbf{S}_j \right) 
+ \text{tr} \left( \boldsymbol{\Sigma}_j \mathbf{S}_j \mathbf{A}_j^{\mathrm{T}} \mathbf{M}_j \mathbf{A}_j \right).
\end{equation}

\subsection{Problem Formulation}
In this paper, we aim to maximize the sum achievable rate of all GUs by jointly optimizing the position of the MAs and the transmit beamforming of the BS. The optimization problem is formulated as
\begin{subequations} 
\begin{flalign}
 \textbf{P1}:\ & \max_{ \mathbf{w}_{n},\mathbf{w}_{j},\mathbf{R}_{0}, \mathbf{s}_m } \quad \sum_{n=1}^N R_n \label{p1a}  \\
 {\rm{s.t.}}  \quad &\mathbf{\bar{a}}_k^{\mathsf{H}} \left( 
\sum_{j=1}^{J} \mathbf{w}_j \mathbf{w}_j^{\mathsf{H}} + 
\sum_{n=1}^{N} \mathbf{w}_n \mathbf{w}_n^{\mathsf{H}} + 
\mathbf{R}_0 
\right) \mathbf{\bar{a}}_k \geq d_k^{2} \Gamma, k \in \mathcal{K}, \label{p1b} \\
 & R_{j} \ge \eta_j + \frac{n_1}{2} \log \left( 1 + \frac{n_1 \left( \det(\mathbf{N}_j \mathbf{M}_j) \right)^{\frac{1}{n_1}}}{\bar{l}_j - \bar{l}_{j,\min}} \right), j \in \mathcal{J}, \label{p1c} \\
 & \sum_{n= 1}^{N} \Vert \mathbf{w}_{n} \Vert^2 + \sum_{j= 1}^{J} \Vert \mathbf{w}_{j} \Vert^2 + \text{tr}(\mathbf{R}_{0}) \le P_{\max}, \label{p1d}  \\
& \left\| \boldsymbol{s}_m - \boldsymbol{s}_{m'} \right\| \geq d_{\min}, \forall m \neq m', \, m, m' \in \mathcal{M}, \label{p1e} \\
 & \boldsymbol{s}_m \in \mathcal{C}_t,  \forall m \in \mathcal{M}, \label{p1f} 
\end{flalign} 
\end{subequations}
where $P_{\max}$ denotes the maximum transmit power of the BS, and $d_{\min}$ represents the minimum distance between any pair of the MAs.
Constraints~\eqref{p1b} and~\eqref{p1c} ensure the QoS requirements for target and the CAVs, respectively.
Constraint~\eqref{p1d} imposes the transmit power limit at the BS.
Constraint~\eqref{p1e} ensures a minimum separation distance between adjacent MAs to avoid collisions or excessive interference.
Finally, constraint~\eqref{p1f} models the mobility limitations of the MAs, such as maximum speed or feasible movement directions.

The inherent non-convexity and highly coupled variables of problem \textbf{P1}, particularly the quadratic inequality, the control QoS bound, and the minimum distance constraint, makes problem \textbf{P1} intractable by conventional solvers. Consequently, developing specialized approaches is necessary to find efficient suboptimal solutions.
To address these challenges, we propose an AO-based algorithm that jointly optimizes the position of the MAs and the transmit beamforming of the BS.

\section{Proposed Solution}

We adopt an AO-based algorithm that decouples the original problem \textbf{P1} into two sub-problems: one focusing on the optimization of the position of the MAs, and the other on the design of the transmit beamforming of the BS. These sub-problems are then solved alternately in an iterative manner.

\subsection{MA Position Optimization}
Given the transmit beamforming $\mathbf{w}_{n},\mathbf{w}_{j},\mathbf{R}_{0}$,
the problem can be simplified as the sub-problem of optimizing the MA position $\mathbf{s}_m$, which is reformulated as
\begin{subequations} 
\begin{flalign}
 \textbf{P2}:\ & \max_{ \mathbf{s}_m } \quad \sum_{n=1}^N R_n \label{p2a}  \\
 {\rm{s.t.}}  \quad &\eqref{p1b},\eqref{p1c},\eqref{p1e},\eqref{p1f}.
\end{flalign} 
\end{subequations}
Clearly, the objective function in problem \textbf{P2} is non-convex with respect to the MA positions $\mathbf{s}_m$, making the optimization problem challenging to solve analytically. To address the MA position sub-problem, we employ the particle swarm optimization (PSO) algorithm.

The algorithm begins by initializing $P$ particles, collected in the set $\mathcal{P} = \{\mathbf{r}_1, \mathbf{r}_2, \ldots, \mathbf{r}_P\}$, where each $\mathbf{r}_p$ represents a candidate solution in the search space. The position of each particle is updated iteratively over $T$ maximum iterations. Let the superscript $i$ denote the iteration index, with $i = 1, 2, \ldots, T$. At the $i$-th iteration, the position of the $p$-th particle is denoted by $\mathbf{r}^i_p = \left( \mathbf{s}^i_{1,p}, \mathbf{s}^i_{2,p}, \ldots, \mathbf{s}^i_{M,p} \right)$,
where $\mathbf{s}^i_{m,p}$ represents the spatial position of the $m$-th MA as determined by the $p$-th particle at iteration $i$. Each particle adjusts its position based on its personal best experience and the global best solution found by the swarm, guiding the search toward high-quality feasible solutions.

During each iteration, the position of each particle is updated by leveraging both its individual best experience, denoted by $\tilde{\mathbf{r}}_{i,p}^{\mathrm{best}}$, and the global best solution discovered by the swarm, denoted by $\tilde{\mathbf{r}}_{\mathrm{gbest}}$. 
The personal best position $\tilde{\mathbf{r}}_{i,p}^{\mathrm{best}}$ is given by $\tilde{\mathbf{r}}_{i,p}^{\mathrm{best}} = \arg \max_{\mathbf{r} \in \mathcal{R}_{i,p}} \mathcal{F}_{\mathrm{fit}}(\mathbf{r})$, where $\mathcal{R}_{i,p}$ represents the set of all positions explored by the $p$-th particle up to the $i$-th iteration, and $\mathcal{F}_{\mathrm{fit}}(\cdot)$ is the fitness function that jointly accounts for the objective and constraints of problem \textbf{P2}. Specifically, the fitness function is formulated as
\begin{equation}
\mathcal{F}_{\mathrm{fit}}(\mathbf{r}_{p}^{i}) = \sum_{n=1}^N R_n(\mathbf{r}_{p}^{i}) - \mu_1 \mathcal{P}_{\rm{sensing}} - \mu_2 \mathcal{P}_{\rm{control}} - \mu_3 \mathcal{P}_{\rm{MA}},
\end{equation}
where $\mu_1$, $\mu_2$, and $\mu_3$ are positive penalty weights associated with constraint violations \eqref{p1b}, \eqref{p1c}, and \eqref{p1e}, respectively.
The penalty $\mathcal{P}_{\rm{sensing}}$, corresponding to the constraint \eqref{p1b}, and $\mathcal{P}_{\rm{control}}$ corresponding to the constraint \eqref{p1c}, are defined respectively as
\begin{equation} 
\mathcal{P}_{\rm{sensing}} = \begin{cases} 
1,\ \text{If constraint \eqref{p1b} is violated},  \\
0,\ \text{Otherwise}.
\end{cases}
\end{equation}
\begin{equation} 
\mathcal{P}_{\rm{control}} = \begin{cases} 
1,\ \text{If constraint \eqref{p1c} is violated},  \\
0,\ \text{Otherwise}.
\end{cases}
\end{equation}
The penalty $\mathcal{P}_{\rm{MA}}$ related to the constraint \eqref{p1e} is given by
\begin{align} \small
\mathcal{P}_{\rm{control}} = \bigg\{ \left( \mathbf{s}^i_{m,p}, \mathbf{s}^i_{m',p} \right)  \bigg|  &
\left\| \mathbf{s}^i_{m,p} - \mathbf{s}^i_{m',p} \right\| < d_{\min}, \nonumber \\ 
& \forall m \neq m', \, m, m' \in \mathcal{M} \bigg\}, 
\end{align}

Then, the swarm’s global best solution $\tilde{\mathbf{r}}_{\mathrm{gbest}}$ is determined as $\tilde{\mathbf{r}}_{\mathrm{gbest}} = \arg \max_{p \in \{1, \ldots, P\}} \mathcal{F}_{\mathrm{fit}}\left( \tilde{\mathbf{r}}_{i,p}^{\mathrm{best}} \right)$,
i.e., the best individual solution across all particles.

Using these guiding positions, the velocity of the $p$-th particle is updated at the $i$-th iteration as follows
\begin{equation}
\tilde{\mathbf{v}}_{p}^{i+1} = \omega \tilde{\mathbf{v}}_{p}^{i}
+ c_1 \tau_1 \left( \tilde{\mathbf{r}}_{i,p}^{\mathrm{best}} - \mathbf{r}_p^{i} \right)
+ c_2 \tau_2 \left( \tilde{\mathbf{r}}_{\mathrm{gbest}} - \mathbf{r}_p^{i} \right),
\end{equation}
where $c_1$ and $c_2$ are cognitive and social learning coefficients, respectively. The random variables $\tau_1, \tau_2 \sim \mathcal{U}(0,1)$ introduce stochasticity to enhance exploration and help escape local optima. The inertia weight $\omega$ balances exploration and exploitation and is dynamically adjusted over iterations according to
\begin{equation}
\omega = \omega_{\max} - \frac{(\omega_{\max} - \omega_{\min}) i}{T},
\end{equation}
where $\omega_{\max}$ and $\omega_{\min}$ denote the initial and final inertia weights, respectively.

Finally, the position of the $p$-th particle is updated as
\begin{equation}
\mathbf{r}_{p}^{i+1} =
\min \left( \max \left( \left(
    \mathbf{r}_{p}^{i} + \pi \tilde{\mathbf{v}}_{p}^{i+1}
    \right), 0 \right), D \right),
\end{equation}
where $\pi$ is a step-size scaling factor that controls the magnitude of position updates, thereby balancing global exploration and local refinement.

\subsection{Transmit Beamforming Optimization}
Given the MA position $\mathbf{s}_m$,
the problem can be simplified as the sub-problem of optimizing the transmit beamforming $\mathbf{w}_{n},\mathbf{w}_{j},\mathbf{R}_{0}$, which can be reformulated as
\begin{subequations} 
\begin{flalign}
 \textbf{P3}:\ & \max_{ \mathbf{w}_{n},\mathbf{w}_{j},\mathbf{R}_{0} } \quad \sum_{n=1}^N R_n \label{p2a}  \\
 {\rm{s.t.}}  \quad &\eqref{p1b},\eqref{p1c},\eqref{p1d}.
\end{flalign} 
\end{subequations}
We can observe that the objective function in problem \textbf{P3} is non-convex with respect to the transmit beamforming $\mathbf{w}_{n},\mathbf{w}_{j},\mathbf{R}_{0}$, making the optimization problem challenging to solve analytically.
Furthermore, the optimization involves a mixture of vector and matrix variables, which introduces additional complexity due to the inherent coupling and non-convex structure of the problem. To facilitate a more tractable formulation, we apply the semidefinite relaxation (SDR) by introducing rank-constrained positive semidefinite matrix variables. Specifically, we define $\mathbf{W}_{n} = \mathbf{w}_{n} \mathbf{w}_{n}^{\mathsf{H}}$, $\mathbf{W}_{j} = \mathbf{w}_{j} \mathbf{w}_{j}^{\mathsf{H}}$, $\mathbf{H}_n = \mathbf{h}_n \mathbf{h}_n^{\mathsf{H}}$, and $\mathbf{H}_j = \mathbf{h}_j \mathbf{h}_j^{\mathsf{H}}$.
According to \eqref{rate}, the achievable rate of the $n$-th GU can be expressed as $R_n =  \log_2 \left( \text{Tr}(\mathbf{W}_{n} \mathbf{H}_n) + E_n \right) - \log_2(E_n)$, where 
$E_n = \sum_{i \neq n}^N \text{Tr}(\mathbf{W}_{i} \mathbf{H}_n) + \text{Tr}(\mathbf{R}_{0} \mathbf{H}_n) + \sum_{j=1}^J \text{Tr}(\mathbf{W}_{j} \mathbf{H}_n) + \sigma^2$.

Hence, we approximate the objective function by the successive convex approximation (SCA), where the $l$-th iteration of the local point of $\mathbf{W}_{n}, \mathbf{W}_{j}, \mathbf{R}_{0}$ is $\mathbf{W}^{(l-1)}_{n}, \mathbf{W}^{(l-1)}_{j}, \mathbf{R}^{(l-1)}_{0}$. The objective $R_n$ under the $l$-th iteration can be given by
{\small \begin{align}
 \hat{R}_n = & \log_2 \left( \text{Tr}(\mathbf{W}_{n} \mathbf{H}_n) + E_n \right) - \log_2(E^{(l-1)}_n) \nonumber \\
 & - \frac{\mathbf{H}_n}{\ln 2 E_n^{(l-1)}} \bigg( 
\sum_{i \ne n}^{N} \text{Tr} \big( \mathbf{H}_n (\mathbf{W}_{i} - \mathbf{W}_{i}^{(l-1)}) \big) \nonumber \\
& + \sum_{j=1}^{J} \text{Tr} \big( \mathbf{H}_n (\mathbf{W}_{j} - \mathbf{W}_{j}^{(l-1)}) \big)+ \text{Tr} \big( \mathbf{H}_n (\mathbf{R}_{0} - \mathbf{R}_{0}^{(l-1)}) \big) \bigg),
\end{align}}where $E^{(l-1)}_n = \sum_{i \neq n}^N \text{Tr}(\mathbf{W}^{(l-1)}_{i} \mathbf{H}_n) + \text{Tr}(\mathbf{R}^{(l-1)}_{0} \mathbf{H}_n) + \sum_{j=1}^J \text{Tr}(\mathbf{W}^{(l-1)}_{j} \mathbf{H}_n) + \sigma^2$.

Similarly, as for \eqref{control_rate}, we have
{\small \begin{align}
 \hat{R}_j = & \log_2 \left( \text{Tr}(\mathbf{W}_{j} \mathbf{H}_j) + F_j \right) - \log_2(F^{(l-1)}_j) \nonumber \\
 & - \frac{\mathbf{H}_j}{\ln 2 F_j^{(l-1)}} \bigg( 
\sum_{i \ne j}^{J} \text{Tr} \big( \mathbf{H}_j (\mathbf{W}_{i} - \mathbf{W}_{i}^{(l-1)}) \big) \nonumber \\
& + \sum_{n=1}^{N} \text{Tr} \big( \mathbf{H}_j (\mathbf{W}_{n} - \mathbf{W}_{n}^{(l-1)}) \big)+ \text{Tr} \big( \mathbf{H}_j (\mathbf{R}_{0} - \mathbf{R}_{0}^{(l-1)}) \big) \bigg),
\end{align}}where $F^{(l-1)}_j = \sum_{i \neq j}^J \text{Tr}(\mathbf{W}^{(l-1)}_{i} \mathbf{H}_j) + \text{Tr}(\mathbf{R}^{(l-1)}_{0} \mathbf{H}_j) + \sum_{n=1}^N \text{Tr}(\mathbf{W}^{(l-1)}_{n} \mathbf{H}_j) + \sigma^2$.

Since the right-hand side of \eqref{p1c} is a decreasing function of $\bar{l}_j$, and the value of $\bar{l}_j$ will be specified by the manually determined training variables in the subsequent simulations, the constraint \eqref{p1c} is convex.  
Moreover, the constraint \eqref{p1d} can be rewritten as
\begin{equation}
\sum_{n= 1}^{N} \text{Tr}(\mathbf{W}_{n}) + \sum_{j= 1}^{J} \text{Tr}(\mathbf{W}_{j}) + \text{tr}(\mathbf{R}_{0}) \le P_{\max}.
\end{equation}

Hence, the problem \textbf{P3} can be approximated as problem \textbf{P3$.l$} in the $l$-th iteration, which is given by
\begin{subequations} 
\begin{flalign}
 \textbf{P3}.l:\ & \max_{ \mathbf{W}_{n},\mathbf{W}_{j},\mathbf{R}_{0} } \quad \sum_{n=1}^N \hat{R}_n \label{p3a}  \\
 {\rm{s.t.}}  \quad &\mathbf{\bar{a}}_k^{\mathsf{H}} \left( 
\sum_{j=1}^{J} \mathbf{W}_j  + 
\sum_{n=1}^{N} \mathbf{W}_n + 
\mathbf{R}_0 
\right) \mathbf{\bar{a}}_k \geq d_k^{2} \Gamma, k \in \mathcal{K},  \\
 & \hat{R}_{j} \ge \eta_j + \frac{n_1}{2} \log \left( 1 + \frac{n_1 \left( \det(\mathbf{N}_j \mathbf{M}_j) \right)^{\frac{1}{n_1}}}{\bar{l}_j - \bar{l}_{j,\min}} \right), j \in \mathcal{J},  \\
 &\sum_{n= 1}^{N} \text{Tr}(\mathbf{W}_{n}) + \sum_{j= 1}^{J} \text{Tr}(\mathbf{W}_{j}) + \text{tr}(\mathbf{R}_{0}) \le P_{\max}, \\
 & \text{rank}(\mathbf{W}_j) = 1, j \in \mathcal{J}, \label{p3b} \\
  & \text{rank}(\mathbf{W}_n) = 1, n \in \mathcal{N}. \label{p3c}
\end{flalign} 
\end{subequations}
Clearly, constraints \eqref{p3b} and \eqref{p3c} remains non-convex. To resolve this, we relax the rank constraints, transforming problem (\textbf{P3$.l$}) into the convex problem (\textbf{RP3$.l$}). Although the relaxed solution may not yield rank-one matrices, an optimal rank-one solution $\mathbf{W}_n^{*}, \mathbf{W}_j^{*}$ always exists, as shown in prior work \cite{10879807}. The proof is omitted due to space constraints. The procedure of the AO-based algorithm is summarized in Algorithm \ref{Alg}.

\begin{algorithm} \footnotesize
	\caption{AO for Solving \textbf{P1}}
	\label{Alg}
	\begin{algorithmic}
		\REQUIRE {An initial feasible point $\mathbf{w}_{n}^{(0)}$, $\mathbf{w}_{j}^{(0)}$, $\mathbf{R}_{0}^{(0)}$, and $\mathbf{s}_{m}^{(0)}$;}\\
		\textbf{Initialize:} the iteration number $o=1$;
		the precision threshold $\epsilon^*$;
            the maximum number of iterations $o_{\rm{max}}$;\\	
		\REPEAT
		\STATE Given $\{ \mathbf{w}_{n}^{(o-1)}$, $\mathbf{w}_{j}^{(o-1)}$, $\mathbf{R}_{0}^{(o-1)} \}$ to solve the problem \textbf{P2}, get the solution $\{ \mathbf{s}_{m}^{(o)} \}$;\\
            \STATE Given $\{ \mathbf{s}_{m}^{(o)} \}$ to solve the problem \textbf{P3}.$l$, get the solution $\{ \mathbf{w}_{n}^{(o)}$, $\mathbf{w}_{j}^{(o)}$, $\mathbf{R}_{0}^{(o)} \}$;\\
		\STATE Set $o=o+1$; \\
		\UNTIL {the gain of objective $\leq \epsilon^*$ or $o = o_{\max}$};\\
		\ENSURE {$\mathbf{w}_{n}^{*}$, $\mathbf{w}_{j}^{*}$, $\mathbf{R}_{0}^{*}$, and $\mathbf{s}_{m}^{*}$.}\\
	\end{algorithmic}
\end{algorithm}

\subsection{ Convergence and Complexity Analysis}
Due to the boundedness of the objective function and the compactness of the feasible set defined by the constraints on all optimization variables, the sequence of the objective values obtained by the Algorithm \ref{Alg} is monotonically non-decreasing and bounded above. Consequently, according to the monotone convergence theorem, the Algorithm \ref{Alg} is guaranteed to converge to a stationary solution.

The computational complexity for solving problem \textbf{P2} is given by $\mathcal{O}(T P^2),$
while the computational complexity for solving Problem \textbf{P3}$.l$ via the interior-point method is
\(
\mathcal{O}\left(( M^2 (  N+J+1 ) )^{3.5} \log\left(\epsilon^{-1}\right)\right),
\)
where \(\epsilon\) denotes the solution accuracy.
As a result, the total computational complexity for solving the optimization problem \textbf{P1} is given by
\(
\mathcal{O}\left( o_{\max} \left(  T P^2 + ( M^2 (  N+J+1 ) )^{3.5} \log\left(\epsilon^{-1}\right) \right)    \right).
\)

\section{Numerical Results}
In this section, we present numerical results to demonstrate the effectiveness of the proposed  AO-based algorithm in the ISSC system, where $3$ GUs, $2$ CAVs, and $3$ targets are assumed to be deployed within a square area of $500\,\mathrm{m} \times 500\,\mathrm{m}$. The BS is located at $[100\ \mathrm{m}, 100\ \mathrm{m}]$, with a height of $10\,\mathrm{m}$. The GUs, the CAVs, and the targets are randomly distributed within the designated area. The antenna array consists of $M = 8$ elements arranged within a $10\,\lambda \times 10\,\lambda$ area, with the \(d_{\min}\) of $0.5\,\lambda$. 
The sensing threshold $\Gamma$ is set to $-15\,\mathrm{dBm}$, and the minimum achievable LQR cost $\bar{l}_{j,\min}$ is $2.8304$. The maximum transmit power $P_{\max}$ is $50\ \mathrm{dBm}$. The reference channel gain $\beta_0$ is $-60\,\mathrm{dB}$, the path-loss exponent \(\alpha\) is $2$, and the noise power $\sigma_n^2=\sigma_j^2=-100\,\mathrm{dBm}$. The Rician factor $K_n$ = $K_j$ = $31.3$. For the PSO algorithm, the maximum number of iterations \( T \) is $300$, the swarm size \( P \) is $200$, the  \( \omega_{\min} \) is $0.4$ and the \( \omega_{\max} \) is $0.9$. The learning coefficients are set as \( c_{1} \) = \( c_{2} \) = $1.5$  and the penalty weights are chosen as \( \mu_{1} \) = \( \mu_{2} \) = \( \mu_{3} \) = $100$. The proposed algorithm is compared with the following benchmark algorithms: Random antenna position (RAP) scheme: 1) The antenna positions are randomly initialized and remain unchanged during the entire optimization process.
2) Fixed antenna position (FAP) scheme: The antenna positions are pre-determined and kept fixed.

\begin{figure}[ht]
    \centering \includegraphics[width=0.35\textwidth]{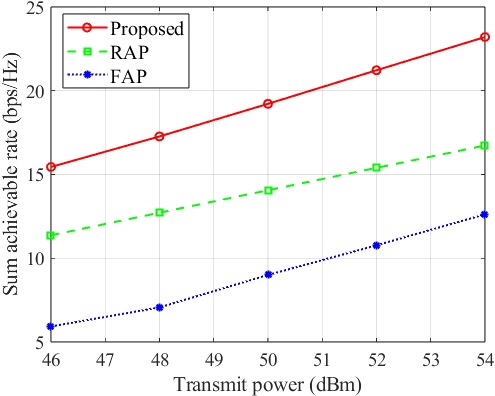}
    \caption{The sum achievable rate versus the transmit power threshold of the BS.}
    \label{fig:pk_img}
\end{figure}
\begin{figure}[ht]
    \centering \includegraphics[width=0.35\textwidth]{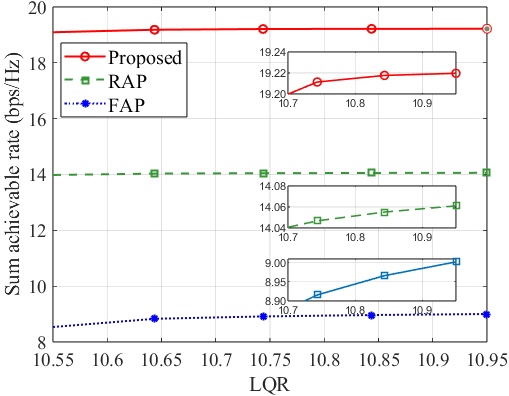}
    \caption{The sum achievable rate versus the LQR cost values.}
    \label{fig:LQR_img}
\end{figure}

As shown in Fig.~\ref{fig:pk_img}, the sum achievable rate versus the transmit power of the BS is presented.
It can be clearly observed that the proposed AO-based scheme consistently outperforms both benchmark schemes in terms of the sum achievable rate with the varying transmit power of the BS.
The reason is that, the optimized APV of the proposed AO-based scheme significantly enhances the BS-GU link quality by delivering higher SINR, consequently outperforming the benchmark schemes with higher sum achievable rates.
In addition, the sum achievable rate of all GUs increases with the transmit power of the BS, as the higher transmit power leads to a higher SINR at the GUs.

Fig.~\ref{fig:LQR_img} shows the sum achievable rate versus different LQR cost values.
As the LQR cost value increases, the sum achievable rate also shows an mildly increasing trend. Notably, the proposed scheme consistently outperforms both the RAP scheme and the FAP scheme across all LQR cost values.
The reason is that, the performance gain is attributed to the effectiveness of the joint optimization of the beamforming of the BS and the APV of the MAs, which enables better adaptation to the positions of the GUs for achieving better communication rates while meeting the requirements of the sensing beampattern gain towards the targtes, and the control stability and performance for the CAVs.
Notably, the RAP scheme consistently outperforms the FAP scheme, since the randomness in antenna positioning introduces spatial diversity, which can potentially lead to more favorable channel conditions and better beamforming opportunities, even without dynamic reconfiguration.

\section{Conclusion}
This paper investigates a novel ISCC system over the LAWN, where the sum achievable rate of all GUs is maximized while ensure the QoS requirements of the sensing beam-pattern gain towards the targets and the control stability and performance for the CAVs.
We propose an AO-based algorithm to solve the non-convex optimization problem efficiently. Then, we analysis the convergence behavior and the computational complexity of the proposed algorithm.
The effectiveness of the proposed algorithm is validated through the numerical results compared with two benchmarks.


\bibliographystyle{ieeetr}
\bibliography{myref}
\end{document}